\documentclass[pra]{revtex4}
 \usepackage{amssymb} \usepackage{graphicx}

\begin{document}
 \title{Generalized Chern-Simons Modified Gravity in First-Order Formalism}
 \author{\"Umit Ertem}
 \email{umitertemm@gmail.com}
\author{\"{O}zg\"{u}r A\c{c}{\i}k}
\email{ozacik@science.ankara.edu.tr}
\address{Department of Physics,
Ankara University, Faculty of Sciences, 06100, Tando\u gan-Ankara,
Turkey\\}

\date{\today}

\begin{abstract}
We propose a generalization of Chern-Simons (CS) modified gravity in
first-order formalism. CS modified gravity action has a term that
comes from the chiral anomaly which is Pontryagin invariant.
First-order CS modified gravity is a torsional theory and in a
space-time with torsion the chiral anomaly includes a torsional
topological term called Nieh-Yan invariant. We generalize the CS
modified gravity by adding the Nieh-Yan term to the action and find
the effective theory. We compare the generalized theory with the
first-order CS modified gravity and comment on the similarities and
differences.
\end{abstract}


\maketitle

\section{Introduction}

The modifications of General Relativity (GR) are widely considered
in the literature to find the true high energy limit of it. One way
of modification is adding some topological terms to the
Einstein-Hilbert Lagrangian. In four dimensions the topological
invariants related to the curvature are Pontryagin and Euler
classes. These are added to the action with associated topological
parameters which are considered as coupling constants \cite{Deser
Duff Isham, Freidel Minic Takeuchi}. Since the Pontryagin class is a
locally exact form and can be written in terms of the Chern-Simons
(CS) form, in the case of constant coupling it can be written as a
boundary term and the equations of motion do not change. However,
the boundary terms change the canonical variables in the Hamiltonian
formalism of the theory and can lead to different ways in canonical
quantization of gravity \cite{Ashtekar}. Another topological
modification is adding the Holst term to the action with constant
coupling parameter called the Immirzi parameter \cite{Holst,
Immirzi, Fatibene Ferraris Francaviglia}. Although the Holst term is
not a topological invariant, it can be considered as the 'half' of a
torsional topological invariant which is mentioned below.

By adding more external fields to the theory one can reach another
type of modification. One way of doing this is promoting the
coupling constant of a topological term to a field. Adding the
Pontryagin term to the action by multiplying a space-time dependent
coupling scalar field leads to the CS modification of GR
\cite{Jackiw Pi}. In CS modified gravity the field equations include
an extra term which is a Cotton-like tensor that is formed by Ricci
tensor and the dual of the Riemann tensor. So, the field equations
are higher than second order in derivatives. But, this does not
prevent the Schwarzschild, Robertson-Walker and gravitational wave
space-times from being solutions of CS modified gravity \cite{Jackiw
Pi}. However, the theory has a constraint resulting from the field
equations that are found from varying the external scalar field.
This constraint requires the vanishing of the Pontryagin term. For a
recent review of CS modified gravity, see \cite{Alexander Yunes
rev}. Different aspects of it are also investigated in the
literature \cite{Grumiller Mann McNees, Smith, Konno, Yunes
Pretorius}.

Recently, the first-order formalism of the CS modified gravity was
considered by several people \cite{Botta Cantcheff, Alexander
Yunes}. The main motivation for this was that treating the curvature
of the underlying manifold as a gauge curvature of the connection
that can be considered as a gauge variable. In pure first-order
theory the equations of motion leads to zero torsion and the theory
reduces to Einstein field equations. However, in the presence of
matter with spin, the theory leads to non-zero torsion. CS modified
gravity is written in first-order formalism as adding the Pontryagin
term with a coupling scalar field to the first-order action.
Although the equations of motion found from varying the co-frame
field leads to the same field equations as in pure theory, the
equations of motion found from varying the connection leads to a
non-zero torsion. So, in first-order formalism, CS modified gravity
is a torsional theory and the torsion is written in terms of the
curvature 2-forms and derivatives of the coupling scalar field.

In the presence of torsion, besides the Pontryagin and Euler classes
there is a torsional topological invariant called Nieh-Yan class
\cite{Nieh Yan, Chandia Zanelli}. So, by coupling with an external
scalar field, one can use it for a modification of GR in a torsional
geometry. Similar to the Pontryagin term, Nieh-Yan class can also be
written as an exact form in terms of a CS-like term. This is called
the translational CS form \cite{Hehl McCrea Mielke Neeman}. Since
first-order formulation of CS modified gravity has torsion, in the
sake of generality, one must also use the Nieh-Yan class in the
topological coupling term. By doing this, we reach a torsional
generalization of CS modified gravity. But, in this theory, field
equations will differ from the pure CS modified gravity and torsion
will have an additional term. Moreover, coupling the theory with
Dirac fermions leads to an effective action that includes more
interaction terms than before.

CS modified gravity is also used as a cancelation mechanism for
Green-Schwarz anomaly since the Green-Schwarz mechanism requires the
inclusion of a Pontryagin term to the action \cite{Alexander Gates}.
Existence of anomalies are related to the topological properties of
the background. In torsion-free case the chiral anomaly is
proportional to the Pontryagin class of the manifold. However, in
the presence of torsion, the chiral anomaly has an additional term
which is proportional to the Nieh-Yan class \cite{Chandia Zanelli 2,
Soo, Li, Mielke}. This is another motivation for generalizing the CS
modified gravity to include the Nieh-Yan term. On the other hand,
the SO(4) Nieh-Yan class is included in the SO(5) Pontryagin class
\cite{Chandia Zanelli 2}. By thinking the embedding of SO(4) into
SO(5) and treating gravity as an SO(5) gauge theory, the
modification with an SO(5) Pontryagin term leads to an SO(4)
Nieh-Yan term. The Nieh-Yan invariant is also used in the literature
for some related manners such as Barbero-Immirzi field
\cite{Mercuri, Mercuri Taveras}. In fact, motivation for the
Nieh-Yan modified gravity was appearance of the Nieh-Yan term in
chiral anomaly in the presence of torsion and this motivates to take
the Barbero-Immirzi parameter as a field \cite{Mercuri Taveras,
Taveras Yunes, Gomez Krasnov, Lattanzi Mercuri}. However,
gravitational chiral anomaly contains both Pontryagin and Nieh-Yan
invariants in the presence of torsion. So, the coupling of a scalar
field with these two invariants is a natural generalization for
anomaly cancelation mechanism. This approach also combines the CS
modified gravity and the Nieh-Yan modified gravity with
Barbero-Immirzi field in one conceptual framework.

In this paper, we add the Nieh-Yan term to the Pontryagin term in
the CS modified gravity. The field equations of generalized CS
modified gravity are found. By coupling with fermions, the torsion
and contorsion forms are obtained and the effective action is
written. So, the new interaction terms are found and the physical
interpretations are discussed.

\section{CS Modified Gravity}

In its original version, CS modified gravity is considered in
second-order formalism \cite{Jackiw Pi, Grumiller Mann McNees}. The action is written as a sum of the Einstein-Hilbert term
and the Pontryagin term with a coupling scalar field;

\begin{eqnarray}
S=\kappa\int d^4x\sqrt{-g}(R+\frac{1}{4}\theta ({}^*RR))\nonumber
\end{eqnarray}
where $\kappa=\frac{1}{16\pi G}$, $g$ is the determinant of the
metric, $R$ is the curvature scalar, $\theta$ is the coupling field
and $^*RR$ is the Pontryagin density which is defined by
\begin{eqnarray}
^*RR={^*R}^{abcd}R_{bacd}\nonumber
\end{eqnarray}
and the dual Riemann tensor is
$^*R^{abcd}=\frac{1}{2}\epsilon^{cdef}{R^{ab}}_{ef}$ with
$\epsilon^{abcd}$ is totally antisymmetric Levi-Civita tensor. The
Pontryagin density can be written as a total derivative of the CS
term. Hence, by taking integration by parts and neglecting the
boundary terms, the second term of the action can be written as
follows
\begin{eqnarray}
S_{CS}=-\frac{\kappa}{2}\int d^4x\sqrt{-g}v_aK^a\nonumber
\end{eqnarray}
where $v_a=\nabla_a\theta=\partial_a\theta$ is the CS velocity and
$K^a$ is the CS term;
\begin{eqnarray}
K^a=\epsilon^{abcd}{\Gamma_{bf}}^e(\frac{1}{2}{R_{cde}}^f-\frac{1}{3}{\Gamma_{ce}}^k{\Gamma_{dk}}^f)\nonumber
\end{eqnarray}
which satisfies $\nabla_aK^a={^*R}R/2$ with $\nabla_a$ is the
covariant derivative with respect to the Levi-Civita connection
${\Gamma_{ab}}^c$ \cite{Jackiw Pi}. The quantity $K^a$ can only be
defined locally, because the connection terms which are used in the
definition of $K^a$ can not globally defined in general.

The field equations are given by
\begin{eqnarray}
G_{ab}+C_{ab}=0\nonumber
\end{eqnarray}
where $G_{ab}=R_{ab}-\frac{1}{2}g_{ab}R$ is the Einstein tensor and
$C^{ab}$ is the Cotton-like tensor which is defined by
\begin{eqnarray}
C^{ab}&=&-\frac{1}{2\sqrt{-g}}(v_e(\epsilon^{eacd}\nabla_c{R^b}_d+\epsilon^{ebcd}\nabla_c{R^a}_d)\nonumber\\
&&+\nabla_e v_f(^*R^{faeb}+^*R^{fbea})).\nonumber
\end{eqnarray}
On the other hand, variation with respect to the coupling scalar
field $\theta$ yields a constraint equation which is called
Pontryagin constraint;
\begin{eqnarray}
^*RR=0.\nonumber
\end{eqnarray}
This implies that the solutions of CS modified gravity must be of
Petrov types $III$, $N$ and $O$  \cite{Alexander Yunes rev}.
Especially, the Schwarzschild, Robertson-Walker and gravitational wave
space-times are solutions of the CS modified gravity.

\section{First-Order Formalism}

In first-order formalism of gravity, independent field variables are
the orthonormal co-frame basis $e^a$ and the connection 1-forms
$\omega^{ab}$. Torsion and curvature 2-forms are defined from these
quantities respectively as;
\begin{eqnarray}
T^a&=&de^a+{\omega^a}_b\wedge e^b\nonumber\\
R^{ab}&=&d\omega^{ab}+{\omega^a}_c\wedge\omega^{cb}\nonumber
\end{eqnarray}
where $d$ is the exterior derivative. Torsion and curvature 2-forms
can be used in the construction of topological invariants. In four
dimensions, the topological invariants defined from curvature are
Pontryagin and Euler classes. These are written as
\begin{eqnarray}
P&=&\frac{1}{8\pi^2}\int R^{ab}\wedge R_{ab}\\
E&=&\frac{1}{16\pi^2}\int R^{ab}\wedge *R_{ab}
\end{eqnarray}
where $*$ is the Hodge star operator on differential forms. CS
modified gravity in first-order formalism is constructed from
Einstein-Hilbert action and the Pontryagin term coupled with an
external scalar field $\theta$;
\begin{eqnarray}
S&=&S_{EH}+S_{CS}\nonumber\\
&=&\kappa\int R^{ab}\wedge *e_{ab}+\frac{1}{2}\int\theta
R^{ab}\wedge R_{ab}.
\end{eqnarray}
Here $e_{ab}=e_a\wedge e_b$. However, the Pontryagin term can be
written as a boundary term, namely exterior derivative of the CS
form;
\begin{eqnarray}
R^{ab}\wedge R_{ab}=d(\omega^{ab}\wedge
R_{ab}-\frac{1}{3}\omega^{ab}\wedge{\omega_a}^c\wedge\omega_{cb})\nonumber.
\end{eqnarray}
So, the second term in the action can be written as
\begin{eqnarray}
S_{CS}=-\frac{1}{2}\int d\theta\wedge(\omega^{ab}\wedge
R_{ab}-\frac{1}{3}\omega^{ab}\wedge{\omega_a}^c\wedge\omega_{cb}).
\end{eqnarray}
By varying the action with respect to $e^a$ and $\omega^{ab}$, the
field equations in first-order formalism are found as
\begin{eqnarray}
G_a=R^{cb}\wedge*e_{cba}&=&0\nonumber\\
\kappa D*e_{ab}+d\theta\wedge R_{ab}&=&0\nonumber
\end{eqnarray}
where $G_a$ is Einstein 3-forms and $D$ is the covariant exterior
derivative. The second equation requires that the torsion is not
zero. Moreover, variation with respect to the coupling scalar field
leads to a constraint that requires the Pontryagin class to be zero;
\begin{eqnarray}
P=R^{ab}\wedge R_{ab}=0\nonumber.
\end{eqnarray}
Hence, in first-order formalism, CS modified gravity is a torsional
theory and this has some implications in the case of interaction
with fermions.

\subsection{Interaction with Fermions}

In the presence of fermions, the coupled action of CS modified
gravity is
\begin{eqnarray}
S=S_{EH}+S_{CS}+S_D\nonumber
\end{eqnarray}
where the Dirac action is written as
\begin{eqnarray}
S_D=\frac{ik}{2}\int*e_a\wedge(\overline{\psi}\gamma^aD\psi-\overline{D\psi}\gamma^a\psi).
\end{eqnarray}
Here $\psi$ is the fermion field, $\gamma^a$ are gamma matrices and
\begin{eqnarray}
D\psi=d\psi+\omega\psi\nonumber
\end{eqnarray}
where $\omega=(1/4)\omega^{ab}\gamma_a\gamma_b$. In the first field
equations, the only difference is the inclusion of the stress-energy
forms of matter $\tau^a$ which are variations of the Dirac action
with respect to the coframe field $e^a$. The second field equations
are changed as
\begin{eqnarray}
\kappa D*e_{ab}+d\theta\wedge R_{ab}+\frac{k}{4}A^ce_{abc}=0.
\end{eqnarray}
Here $e_{abc}=e_a\wedge e_b\wedge e_c$ and
\begin{eqnarray}
A^c=\bar{\psi}\gamma_5\gamma^c\psi\nonumber
\end{eqnarray}
is the axial current. In this case, the torsion 2-forms are found as
below \cite{Alexander Yunes rev}
\begin{eqnarray}
T_a=\frac{1}{2\kappa}v^b\epsilon_{bapq}R^{pq}+\frac{k}{8\kappa}A^b\epsilon_{bapq}e^{pq}
\end{eqnarray}
where $i_X$ is the interior derivative with respect to the vector field
$X$ and by defining $v^a=i_{X^a}d\theta$ we neglect the terms that
are second order in $v$. On the other hand, contorsion 1-forms are
defined as the difference between the connection 1-forms
$\omega^{ab}$ and the Levi-Civita connection 1-forms $\Gamma^{ab}$,
namely $C^{ab}=\omega^{ab}-\Gamma^{ab}$. So, the torsion can be
written in terms of the contorsion 1-forms as
\begin{eqnarray}
T^a={C^a}_b\wedge e^b\nonumber.
\end{eqnarray}
Hence, the contorsion 1-forms are found from the equation below
\begin{eqnarray}
C^{ab}=\frac{1}{2}(-e^di_{X^a}i_{X^b}T_d-i_{X^b}T^a+i_{X^a}T^b),
\end{eqnarray}
and by using the torsion found above we arrive at \cite{Alexander
Yunes rev}
\begin{eqnarray}
C_{ab}=\frac{1}{8\kappa}v_k{\epsilon_{[a}}^{kqp}R_{bd]qp}e^d+\frac{3k}{8\kappa}A^k\epsilon_{kdba}e^d
\end{eqnarray}
where $R_{abcd}$ are the components of the curvature tensor.

By reinserting the torsion $T^a$ as ${C^a}_b\wedge e^b$ into the action, one can obtain the
effective action, namely this leads to the action that is written as
a sum of two terms $S=S[\Gamma]+S[C]$. The first one is the torsion
free part and the second one includes contorsion-induced interaction
terms. So, the effective action is written as
\begin{eqnarray}
S_{eff}&=&S(\Gamma)+\kappa\int C_{ac}\wedge {C^c}_b\wedge*e^{ab}\nonumber\\
&&+\int\theta[R_{ab}(\Gamma)\wedge
D_{\Gamma}C^{ab}+\frac{1}{2}R_{ab}(\Gamma)\wedge {C^a}_c\wedge
C^{cb}\nonumber\\
&&+D_{\Gamma}C_{ab}\wedge{C^a}_c\wedge
C^{cb}+\frac{1}{2}C_{ad}\wedge C^{db}\wedge{C^a}_c\wedge
{C^{c}}_b\nonumber\\
&&+\Gamma_{ac}\wedge{C^c}_b\wedge(R^{ab}(\Gamma)+D_{\Gamma}C^{ab}+{C^a}_d\wedge
C^{db})]\nonumber\\
&&+\frac{ik}{2}\int*e_a\wedge (\bar{\psi}\gamma^a
C\psi-C\bar{\psi}\gamma^a\psi)\nonumber.
\end{eqnarray}
Here $D_\Gamma$ is the covariant exterior derivative with respect to
the torsion-free connection $\Gamma$, $R_{ab}(\Gamma)$ are the
curvature 2-forms that are constructed from $\Gamma$ and
$C=\frac{1}{4}C_{ab}\gamma^a\gamma^b$. This procedure does not change the field equations.

By using the contorsion found above the effective action can be
written as \cite{Alexander Yunes}
\begin{eqnarray}
S_{eff1}&=&S(\Gamma)+\int [\frac{k}{16\kappa}(2A^av^bi_{X_b}P_a-A^av_a{\cal{R}})\nonumber\\
&&+\frac{3k^2}{32\kappa}A^aA_a-\frac{3k^2}{196\kappa^2}\epsilon^{abcd}v_aA_d\partial_cA_b\nonumber\\
&&-\frac{k^3}{256\kappa^3}v_aA^aA^bA_b]*1+O(v^2)
\end{eqnarray}
where $P_a$ are Ricci 1-forms and ${\cal{R}}$ is the curvature
scalar. Here the terms including contractions with connection
1-forms are neglected by assuming the Riemann normal coordinates.
The four-fermion interaction term comes from the first-order
formalism without CS modification. So, the CS modification differs
mainly by the interaction term of two fermions with CS velocity
$v_a$ and curvature characteristics. The other terms are suppressed
by the higher powers of the Newton's constant $G$.

\section{Generalized CS Modified Gravity}

As we have seen, in CS modified gravity coupled with fermions, the
torsion is not zero. However, in the presence of torsion there is
one more topological invariant called Nieh-Yan class related with
torsion;
\begin{eqnarray}
N=\int(T^a\wedge T_a-R_{ab}\wedge e^{ab})
\end{eqnarray}
and this can also be written as a boundary term;
\begin{eqnarray}
T^a\wedge T_a-R_{ab}\wedge e^{ab}=d(e^a\wedge T_a)\nonumber.
\end{eqnarray}
Hence, one can generalize the CS modified gravity action by
including the Nieh-Yan term besides the Pontryagin term. We will see
that this will add a small modification to the terms in the
effective action that are not suppressed by the higher powers of the
Newton's constant;
\begin{eqnarray}
S&=&S_{EH}+S_{CS}+S_{NY}\nonumber\\
&=&\kappa\int R^{ab}\wedge *e_{ab}\\
&&+\frac{1}{2}\int\theta (R^{ab}\wedge
R_{ab}+\frac{2}{l^2}(T^a\wedge T_a-R_{ab}\wedge e^{ab}))\nonumber
\end{eqnarray}
where $l$ is a constant in the units of length. The factor in front
of the Nieh-Yan term is added for writing the action in appropriate
units. As before, the last term in the action can be written as
\begin{eqnarray}
S_{CS}+S_{NY}&=&-\frac{1}{2}\int d\theta\wedge(\omega^{ab}\wedge
R_{ab}-\frac{1}{3}\omega^{ab}\wedge{\omega_a}^c\wedge\omega_{cb}\nonumber\\
&&-\frac{2}{l^2}e^a\wedge T_a).
\end{eqnarray}

To find the field equations of the generalized CS modified gravity,
we take the variations of the action (12). By varying (12) with
respect to $e^a$ and neglecting the boundary terms, one finds that
\begin{eqnarray}
\delta S&=&\kappa\int R^{ab}\wedge*e_{abc}\wedge\delta
e^c-\frac{2}{l^2}\int \delta e_a\wedge d\theta\wedge T^a\nonumber.
\end{eqnarray}
Hence the first field equations are found as
\begin{eqnarray}
\kappa R^{bc}\wedge*{e_{bc}}^a-\frac{2}{l^2}d\theta\wedge
T^a=0\nonumber.
\end{eqnarray}
This differs from the without Nieh-Yan case with the inclusion of
torsion and the derivative of the coupling field. On the other hand,
varying (12) with respect to $\omega^{ab}$ leads to
\begin{eqnarray}
\delta S&=&\kappa\int\delta\omega^{ab}\wedge
D*e_{ab}+\int\delta\omega^{ab}\wedge (d\theta\wedge
R_{ab}\nonumber\\
&&+\frac{1}{l^2}d\theta\wedge e_{ab})\nonumber.
\end{eqnarray}
So, the second field equations are found as below
\begin{eqnarray}
\kappa D*e_{ab}+d\theta\wedge(R_{ab}+\frac{1}{l^2}e_{ab})=0.
\end{eqnarray}
Existence of the last term which did not show up in the previous
case is due to the Nieh-Yan term. In the generalized CS modified
gravity, variation of the action with respect to the coupling scalar
field leads to a constraint again, but this time the constraint
requires that the Pontryagin term and Nieh-Yan term must be equal to
each other;
\begin{eqnarray}
R^{ab}\wedge R_{ab}=\frac{2}{l^2}(T^a\wedge T_a-R_{ab}\wedge
e^{ab}).
\end{eqnarray}
This may seem as a restrictive constraint. However, it has a special
solution in the context of Macdowell-Mansouri formulation of gravity
\cite{MacDowell Mansouri}. In Macdowell-Mansouri gravity, connection
1-forms $\omega^{ab}$ and coframe fields $e^a$ are combined into a
new $SO(4,1)$ (or $SO(3,2)$) connection $A^{IJ}$ for $I,J=0,1,...,4$
and $i,j=0,1,...,3$ as follows;
\begin{eqnarray}
A^{ij}=\omega^{ij}\quad\quad,\quad\quad
A^{i4}=\frac{1}{l}e^i\nonumber.
\end{eqnarray}
So the curvature $F^{IJ}=dA^{IJ}+A^{IK}\wedge {A_K}^J$ is written
as;
\begin{eqnarray}
F^{ij}=R^{ij}-\frac{1}{l^2}e^i\wedge e^j\quad\quad,\quad\quad
F^{i4}=\frac{1}{l}T^i\nonumber
\end{eqnarray}
and the Macdowell-Mansouri action is defined as;
\begin{eqnarray}
S_{MM}=\int F_{IJ}\wedge \star F^{IJ}\nonumber
\end{eqnarray}
where $\star$ is the Hodge star defined on internal indices
$I$,$J$,... For flat Macdowell-Mansouri curvature $F=0$ we have
$T^i=0$ and $R^{ij}=\frac{1}{l^2}e^i\wedge e^j$, namely a constant
curvature spacetime. If we define the cosmological constant
$\Lambda$ as $\Lambda=\frac{3}{l^2}$, then this case corresponds to
de Sitter (or anti-de Sitter) spacetime. This special case is a
solution for the above constraint and this may motivate us for
generalizing the action with a cosmological constant term. Although
the cosmological constant term will have an effect on field
equations, it will not affect torsion or contorsion. However, as we
will see below, the effective terms that come from the Nieh-Yan
invariant will be proportional to cosmological constant. This is an
important difference from the ordinary CS modified case.

Now, we can see the effects of the Nieh-Yan term on the CS modified
gravity coupled with fermions. In the fermion coupled generalized CS
modified gravity, there will be an additional contribution to the
torsion. In this case we have the action
\begin{eqnarray}
S=S_{EH}+S_{CS}+S_{NY}+S_D\nonumber.
\end{eqnarray}
Effect of the inclusion of matter to the action on the first field
equations is only the addition of the stress-energy forms of matter
$\tau^a$;
\begin{eqnarray}
\kappa R^{bc}\wedge*{e_{bc}}^a-\frac{2}{l^2}d\theta\wedge
T^a+\tau^a=0.
\end{eqnarray}
The second field equations also change because of the existence of
matter with spin. In this case equation (14) turns to
\begin{eqnarray}
\kappa
D*e_{ab}+d\theta\wedge(R_{ab}+\frac{1}{l^2}e_{ab})+\frac{k}{4}A^c
e_{cab}=0.
\end{eqnarray}
Equation (17) leads to a non-zero torsion and it is given in terms
of the curvature 2-forms, the CS velocity and the axial current by
\begin{eqnarray}
T_a=\frac{1}{2\kappa}v^b\epsilon_{bapq}(R^{pq}+\frac{1}{l^2}e^{pq})+\frac{k}{8\kappa}A^b\epsilon_{bapq}e^{pq}\quad
\end{eqnarray}
Hence, in the light of (8), the contorsion 1-forms are found as (by
neglecting the terms that are second order in $v$)
\begin{eqnarray}
C_{ab}&=&\frac{1}{8\kappa}v_c{\epsilon_{[a}}^{cqp}R_{bd]qp}e^d+\frac{3}{2\kappa l^2}v^c\epsilon_{cdba}e^d\nonumber\\
&&+\frac{3ik}{8\kappa}A^c\epsilon_{cdba}e^d.
\end{eqnarray}

By reinserting the torsion into the action, one can obtain the
effective action and see the differences from the CS modified
gravity. The extra terms that are originated from the Nieh-Yan term
in the effective action can be written as below (by neglecting the
terms that includes contractions with connection 1-forms);
\begin{eqnarray}
S_{eff2}&=&S_{eff1}+\frac{54k}{\kappa l^2}\int A^av_a*1\nonumber\\
&&+\frac{9}{8}\frac{k}{\kappa^2l^2}\int[A^kv_kR_{abpq}\epsilon^{abpq}\nonumber\\
&&+(A_kv^b+A^bv_k)R_{abpq}\epsilon^{pqka}]*1\nonumber\\
&&-\frac{27}{32}\frac{k^2}{\kappa^3l^2}\int[5\partial_aA^aA^bv_b-\partial_aA_b(A^av^b+A^bv^a)]*1\nonumber
\end{eqnarray}
Hence, only the first extra term is at the order of $G$ and all the
other extra terms are suppressed by at least the square of the
Newton's constant. The main difference between the contributions for
the effective theories of CS modified and generalized CS modified
theories is the term that includes two-fermion and CS velocity
interaction without curvature terms;
\begin{eqnarray}
S_{ext}=54\frac{k}{\kappa l^2}\int A^av_a*1.
\end{eqnarray}
The other differences occur only on the $G^2$ order and can be
neglected.

In fact, the effective action contains a kinetic term and possible
other terms for the coupling field if we do not neglect the $O(v^2)$
terms. This may be compared with the pseudo-scalar degree of freedom
of Nieh-Yan modified gravity \cite{Calgagni Mercuri}. However in
this case, all $O(v^2)$ terms and also the kinetic term appear at
least the order of $G^2$. So, the comparison with the ordinary CS
modified case will not be effected by the kinetic term. On the other
hand, the constraint (15) can be written effectively in terms of
contorsion 1-forms;
\begin{eqnarray}
&&R_{ab}(\Gamma)\wedge[\frac{1}{2}R^{ab}(\Gamma)+D_{\Gamma}C^{ab}+C^{ac}\wedge
{C_c}^{b}]\nonumber\\
&&=\frac{2}{l^2}C_{ab}\wedge {C^a}_d\wedge
e^{bd}-D_{\Gamma}C_{ab}\wedge C^{ac}\wedge
{C_c}^{b}\\
&&-\frac{1}{2}C_{ac}\wedge {C^{c}}_b\wedge C^{ad}\wedge
{C_d}^b\nonumber
\end{eqnarray}
From (19) this gives a complicated dynamical equation for the
coupling field and the geometrical constraint (15) is transformed
effectively to an equation for the dynamical field and fermions
coupled to curvature characteristics.

\section{Conclusion}

In the first-order formalism of CS modified gravity, the non-zero
torsion is unavoidable. So, the main motivation of CS modified
gravity, namely the anomaly cancelation and adding a term to the
action which is the Pontryagin invariant coupled with an external
scalar field, must be considered in torsional space-time. In this
case, the Nieh-Yan term can also be added to the action as a
topological invariant which arises in the chiral anomaly with
torsion.

It is known that in first-order CS modified gravity, the new effect
different from the zero torsion case is the two-fermion interaction
with CS velocity contracted by curvature characteristics. This
effect can arise in large fermion current regions with space-time
having large curvature characteristics such as merging binary
neutron stars and neutron star-black hole systems \cite{Alexander
Yunes}.

By considering Nieh-Yan term in the first-order CS modified gravity,
we have arrived at the generalized CS modified gravity. In this
case, we have all the correction terms that arise in CS modified
gravity and some extra terms resulting from the Nieh-Yan invariant.
The main difference, which is first order in $G$, is the term of
two-fermion interaction with CS velocity without curvature
characteristics. This extra term can have effects in large fermion
current regions which do not need to be high curvature space-time
regions.

\begin{acknowledgments}
This work was supported in part by the Scientific and Technical
Research Council of Turkey (T\"{U}B\.{I}TAK).
\end{acknowledgments}


\end{document}